\documentclass[fleqn,usenatbib]{mnras}

\usepackage{newtxtext,newtxmath}

\usepackage[T1]{fontenc}

\DeclareRobustCommand{\VAN}[3]{#2}
\let\VANthebibliography\thebibliography
\def\thebibliography{\DeclareRobustCommand{\VAN}[3]{##3}\VANthebibliography}

\usepackage{graphicx}	
\usepackage{amsmath}	

\title[NGC 2403 Globular Clusters]{Low-Metallicity Globular Clusters in the Low-Mass Isolated Spiral Galaxy NGC 2403}

\author[D. A. Forbes et al.]
{Duncan A. Forbes,$^{1}$\thanks{E-mail: dforbes@swin.edu.au (DAF)}
Anna Ferr\'e-Mateu,$^{2,1}$
Jonah S. Gannon$^{1}$, Aaron J. Romanowsky$^{3,4,5}$, 
\newauthor 
Jeffrey L. Carlin$^{6}$, Jean P. Brodie$^{1,4}$  and Jacob Day$^3$
\\
$^1$Centre for Astrophysics \& Supercomputing, Swinburne University of Technology, Hawthorn VIC 3122, Australia\\
$^2$Instituto Astrofisica de Canarias, Av. Via Lactea s/n, E38205 La Laguna, Spain\\
$^3$Department of Physics and Astronomy, San Jos\'e State University, San Jos\'e, CA 95192, USA\\
$^4$University of California Observatories, 1156 High St., Santa Cruz, CA 95064, USA\\
$^5$Department of Astronomy and Astrophysics, University of California, Santa Cruz, CA 95064, USA\\
$^{6}$NSF's NOIRLab/Rubin Observatory Project Office, 950 North Cherry Avenue, Tucson, AZ 85719, USA
}

\date{Accepted XXX. Received YYY; in original form ZZZ}

\pubyear{2022}

\begin{document}
\label{firstpage}
\pagerange{\pageref{firstpage}--\pageref{lastpage}}
\maketitle

\begin{abstract}

The globular cluster (GC) systems of low-mass late-type galaxies, such as NGC 2403, have been poorly studied to date. As a low mass galaxy (M$_{\ast}$ = 7 $\times$ 10$^{9}$ M$_{\odot}$), cosmological simulations predict NGC 2403 to contain few, if any, accreted GCs. It is also isolated, with 
a remarkably undisturbed HI disk. 
Based on candidates from the literature, Sloan Digital Sky Survey and 
Hyper Suprime-Cam imaging, we selected several GCs for follow-up spectroscopy using the Keck Cosmic Web Imager. From their radial velocities, and other properties, we identify 8 bona-fide GCs associated with either the inner halo or the disk of this bulgeless galaxy.
A stellar population analysis suggests a wide range of GC ages from shortly after the Big Bang until the present day. We find all of the old GCs to be metal-poor with [Fe/H] $\le$ --1.   
The age--metallicity relation for the observed GCs suggests that they were formed over many Gyr from gas with a low effective yield, similar to that observed in the SMC. Outflows of enriched material may have contributed to the low yield. 
With a total system of $\sim$50 GCs expected, our study is the first step in fully mapping the star cluster history of NGC 2403 in both space and time. 

\end{abstract}




\section{Introduction}

The globular cluster (GC) systems of galaxies is an area of active research 
(see review by \citet{2018RSPSA.47470616F} and references therein). This is particularly true for low-mass galaxies, which have relatively poor GC systems compared to  
giant galaxies with their populous GC systems.
While the Local Group contains some well-known low mass galaxies with modest GC systems, such as M33, the LMC and the SMC, the study of their GC systems is in some ways complicated by their proximity. In the case of M33 its radial velocities of its GCs ($\sim$ -180 km/s) overlap with high velocity stars of the Milky Way; 
the recent work of Larsen et al. (2021) found that several claimed GCs, in previous studies of M33, are actually foreground Milky Way stars. The Magellanic Clouds are of course in the process of merging with the Milky Way and it has been speculated for some time that they may have `lost' GCs to our Galaxy \citep{1995MNRAS.275..429L}. Thus we need to look further afield to improve our understanding of the GC systems of such low mass galaxies. 

While GC systems, in general, include both in-situ and ex-situ (accreted) formed GCs \citep{2018MNRAS.479.4760F}, low mass galaxies are expected to be dominated by in-situ formed GCs e.g. \citet{2010ApJ...725.2312O} and 
\citet{2021arXiv210112216R}. In particular, galaxies with a total halo mass of M$_{h}$  
$<$ 10$^{11}$ M$_{\odot}$ are expected to be dominated by in-situ formed GCs \citep{2019MNRAS.488.5409C}. This is in  contrast to massive galaxies like the Milky Way, the halos of which are dominated by the accreted GCs of disrupted dwarf galaxies  \citep{2020MNRAS.493..847F}. 

NGC~2403 is a low mass (M$_V$ = --19.5, M$_{\ast}$ = 7 $\times$ 10$^{9}$ M$_{\odot}$), late-type (Scd) spiral galaxy in the outskirts of the M81 group at a distance of 3.2 Mpc \citep{2016ApJ...828L...5C}. It may be infalling for the first time into the M81 group \citep{2013ApJ...765..120W} with 
M81 itself lying $\sim$1 Mpc away. 
It has a satellite galaxy DDO~44 (M$_V$ = --12.5), and recently a smaller dwarf, called MADCASH-1 (M$_V$ = --7.7), was discovered by  \citep{2016ApJ...828L...5C}. Although DDO~44 shows signs of interacting \citet{2019ApJ...886..109C}, NGC 2403 itself
reveals a remarkably undisturbed HI disk (\citealp{2008AJ....136.2648D}; \citealp{2013ApJ...765..120W}). This, and its relative isolation, suggests it is a galaxy that has not undergone a strong tidal interaction nor experienced ram pressure stripping. 
Although bulgeless, it does host a nuclear star cluster \citep{2007MNRAS.382.1552L} and it reveals an extended stellar structure at large radii which is either an extension of the thick disk or the presence of a faint stellar halo \citep{2012MNRAS.419.1489B}. It is also an important Cepheid variable calibrator galaxy \citep{2006ApJS..165..108S}. 

NGC 2403 has some global properties similar to those of the bulgeless Local Group galaxy M33 (Scd, M$_V$ = --19.4).  
M33 contains a large population of intermediate-age star clusters, particularly at large galactocentric radii (\citealp{2002ApJ...564..712C}; \citealp{2003AJ....125.3046D};  \citealp{2010ApJ...720.1674S}; \citealp{2014ApJS..211...22F}) with 85\% of them kinematically-associated with the inner halo and 15\% associated with the disk \citep{2002ApJ...564..712C}. 
\citet{2002ApJ...564..712C} found that 
the inner halo GCs in M33 reveal a significant age spread when compared to those of the Milky Way. A key difference is that M33 has signatures of an {\it inner} stellar halo but not an {\it outer} one
\citep{2021arXiv211015773G}.
This may indicate a relatively small contribution from accreted dwarf galaxies in M33's assembly history. 
In this context, we note at least one halo GC (HM33-B) which \citet{2021arXiv211200081L}
argued was accreted on the basis of its anomalous chemical signature.

The GC system of NGC 2403 has been poorly studied to date with very little work on it since the photographic plate study of \citet{1984A&A...130..162B}. In this work they identified 8 candidates as having colours typical of GCs, 6 star clusters with bluer colours and 5 with redder colours than typical Milky Way GCs.  \citet{2007ApJ...664..820D} also identified another half dozen GC candidates in their infrared imaging and estimated their sizes. 
NGC 2403 
does not appear in the catalogue of \citet{2013ApJ...772...82H} listing 
galaxies with known GC systems (it lists M33 to have a total of 50 GCs) and its total number of GCs is unknown. Based on its total halo mass of 2--3$\times$10$^{11}$ M$_{\odot}$
\citep{2020ApJS..247...31L}, and the relation of 
\citet{2020AJ....159...56B}, we predict 
a total GC system of 40--50 members. 
Based on a typical GC specific frequency for spiral galaxies, over a range of luminosity, of S$_N$ $\sim$ 1 (see 
\citealp{2021MNRAS.tmp.2639L}), around 60 GCs are expected. 
As well as having an undefined number of GCs, we are unaware of any published high quality spectra for confirmed GCs in this nearby galaxy. Thus 
the fraction of disk vs halo GCs and whether its halo GCs resemble those of old halo GCs of the Milky Way, or its younger accreted GCs are unknown. The  age-metallicity relation (AMR) of NGC 2403's GC system, which potentially holds important clues on the galaxy's assembly history, is currently undefined.

Here we obtain Keck Cosmic Web Imager (KCWI) spectra for 9 GC candidates around NGC 240 finding 7 to be confirmed GCs. Along with 
one other GC in NGC 2403 observed using HIRES on the Keck telescope, bringing the total to 8, we discuss their stellar population properties and compare them to those of SMC/LMC, M33 and the Milky Way. 
At a distance of 3.2 Mpc \citep{2003AJ....125.3046D}, 1" corresponds to 15 pc. 

\begin{figure}
\begin{center}
\includegraphics[width=0.48\textwidth, trim=0in 0.0in 0in 0.0in, clip]{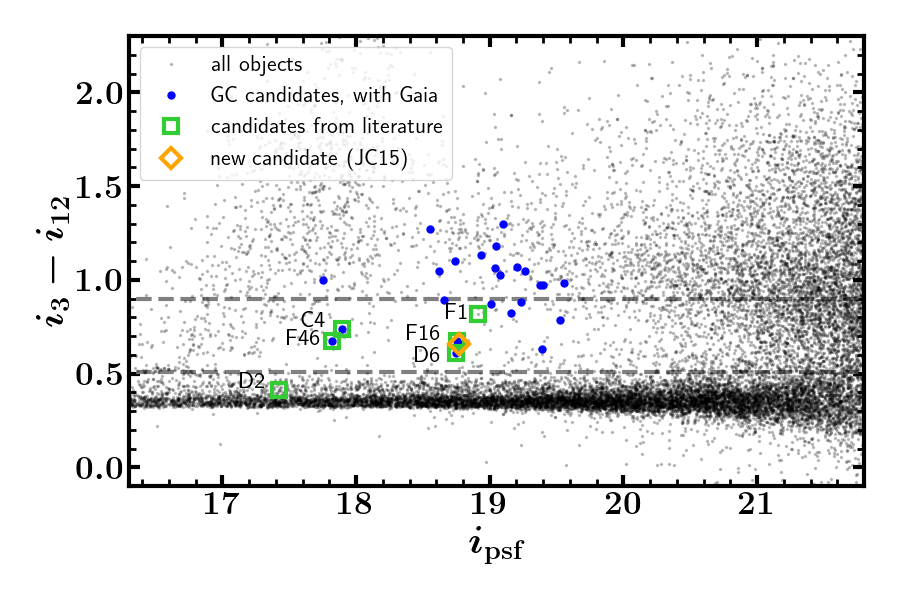}
\caption{Difference between the $i$ band magnitudes measured on a 30s HSC exposure of NGC 2403 with 3-pixel and 12-pixel apertures, $i_3 - i_{12}$, as a function of the i band PSF magnitude. Point sources occupy the locus at $i_3 - i_{12} \sim 0.35$. More extended objects containing excess flux (relative to the PSF) exhibit larger $i_3 - i_{12}$ values, with those between 0.5 and 0.9 (dashed lines) the best candidates for GCs. We highlight as blue dots the two dozen candidate GCs selected based on all of our criteria, open green squares highlight the six candidate GCs with matches to our HSC catalog, and the orange diamond highlights the new candidate, JC15, that we followed up spectroscopically (see section 2.2).}
\label{fig:i3_6}
\end{center}
\end{figure}

\begin{figure}
\begin{center}
\includegraphics[width=0.48\textwidth, trim=0in 0.0in 0in 0.0in, clip]{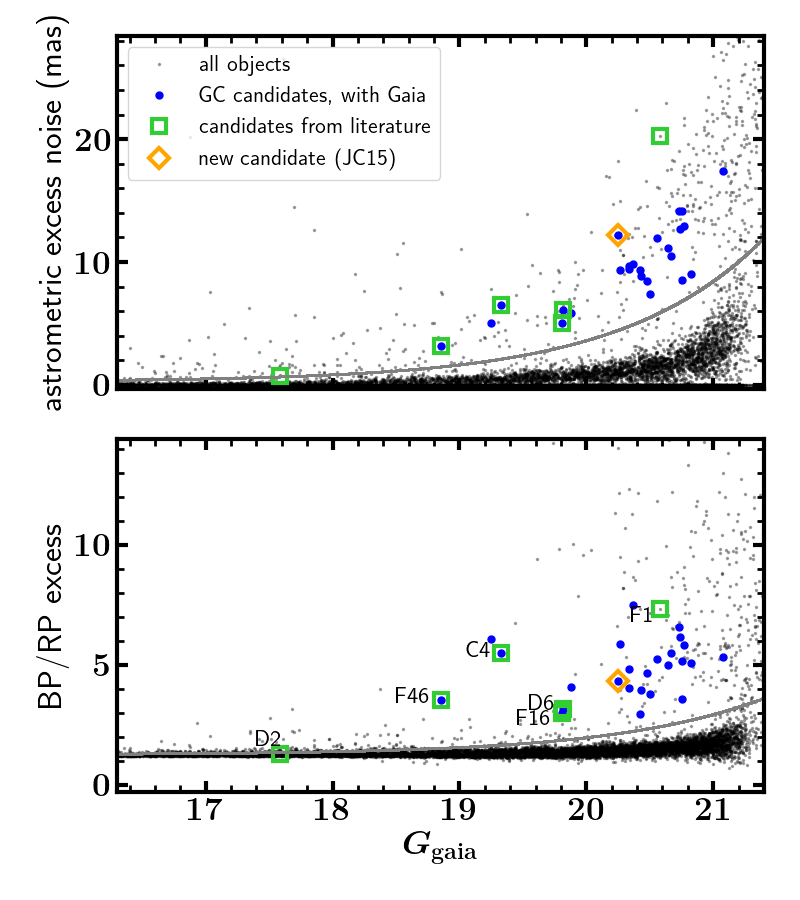}
\caption{Gaia parameters ``astrometric excess noise'' (upper panel) and ``BP/RP excess'' (lower panel) as a function of Gaia G magnitudes for objects in our HSC imaging. The curves in each panel are the GC candidate selection criteria derived by \citet{2021ApJ...914...16H} to select Cen A clusters. Symbols are as in Figure~\ref{fig:i3_6}. }
\label{fig:gaia_excess}
\end{center}
\end{figure}

\begin{table}
	\centering
	\caption{NGC 2403 Globular Cluster Candidate Properties}
	\label{tab:table1}
	\begin{tabular}{lccclrc} 
		\hline
		ID & $g$ & $g-i$ & Source & R$_p$ & Size\\
		& (mag)  & (mag) & & (arcmin) & (pc)\\
		\hline
		C1 & 19.1 & 0.15 & PS & 5.3 & 3.5\\
		C4 &  18.2 & 0.91 & DR15 & 5.3 & 2.0\\
		D2 &  18.0 & 0.68 & DR15 & 8.6 & --\\
		D6 &  19.3 & 0.95 & DR7 & 14.3 & 3.6\\
		F1 &  19.1 & 0.88 & DR7 & 12.4 & 5.0\\
		F16 & 19.5 & 1.01 &  DR15 & 7.9 & 2.2\\
		JD1 &  19.0 & 0.81 & DR7 & 6.7 & --\\
		JD2 &  19.8 & 0.86 & DR15 & 80 & --\\
		JC15 & 19.8 & 1.04 & HSC & 4.2 & --\\
		\hline
		F46 &  18.3 & 0.85 & DR15 & 5.0 & 4.4\\ 
		\hline
	\end{tabular}
	\\
	Notes: Source of photometry, i.e. PanSTARRS, SDSS or Hyper Suprime-Cam imaging. R$_p$ is the projected radius from the centre of NGC 2403 (JD2 lies close to DDO~44), with 1 arcmin equivalent to 900 pc. Sizes for D6, F1 are from \citet{2007ApJ...664..820D}, C1,  C4, F16 from this work and F46 from L21. 
\end{table}

\section{Globular Cluster Candidates}

Our GC candidates for follow-up spectroscopy have been taken from the literature and supplemented by our own data. 

\subsection{Literature Candidates}

From 
\citet{1984A&A...130..162B} we include their objects C1, C4, F1 and F16 and from \citet{2007ApJ...664..820D} we include D2 and D6.
According to \citet{2007ApJ...664..820D}, several GC candidates are partially resolved in their near-IR imaging. They measured 
extended sizes for D2, D6, F1, F16 and F46.
The candidates C1, C4, F16 and F46 are clearly resolved GCs in archival {\it Hubble Space Telescope Advanced Camera for Surveys} (HST/ACS) imaging. We have estimated the effective radius of each of these GCs after subtracting in quadrature the ACS point spread function (based on several; stars in the image). Our radii are listed in Table 1, with F46 from 
\citet{2021arXiv211200081L}. While we find a similar size for F16 (2.2 vs 2.7 pc), we can rule out F46 having a size of 10.4 pc, as found by \citet{2007ApJ...664..820D}, based on the ACS imaging. For D2 they estimated a small but extended size of 1.8 pc. We also examined the other candidates listed in \citet{1984A&A...130..162B} 
with available HST/ACS imaging but concluded that are unlikely to be GCs.



\begin{figure}
	\includegraphics[width=1.08\columnwidth,angle=0]{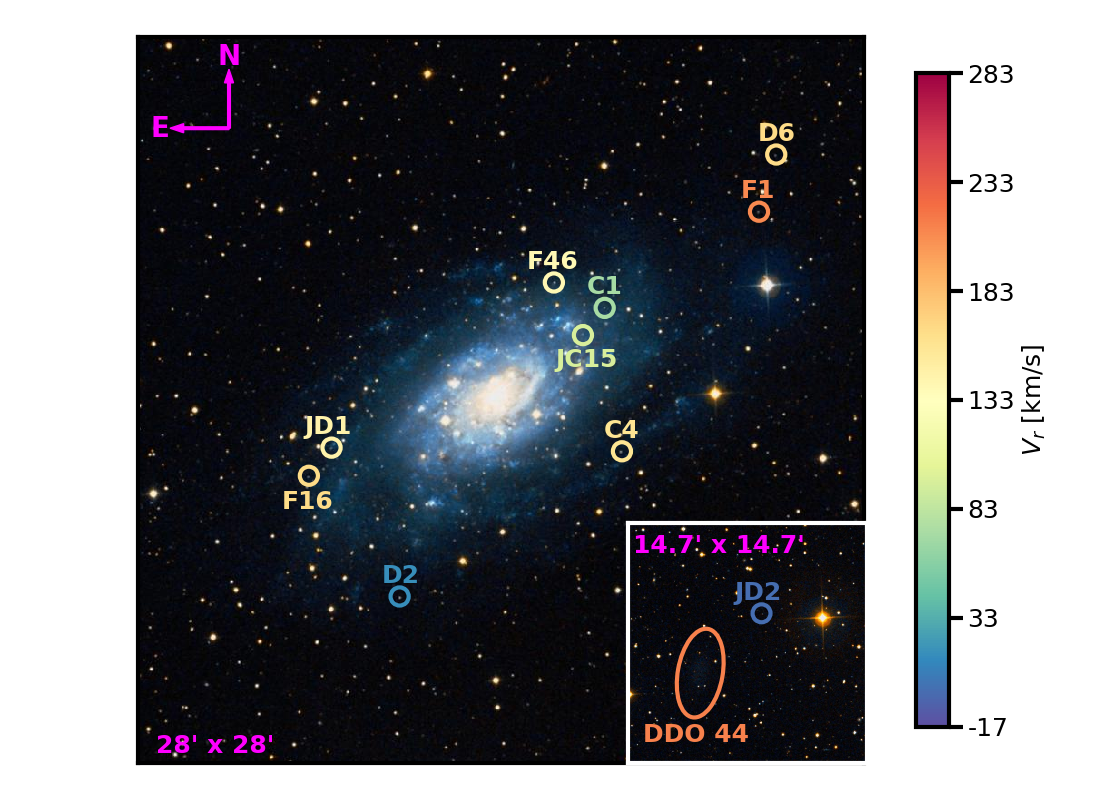}
    \caption{
    A 28' $\times$ 28' colour SDSS image of NGC~2403 taken from \href{http://wikisky.org}{wikisky}. The globular cluster candidates studied in this work are indicated via open circles. They are labelled appropriately and colour coded by their recession velocity (RV). We note that NGC~2403 has an RV of 133 km/s and the receding side of the galaxy stellar/HI disk (i.e. higher RVs) is the SE side. A 14.7' $\times$ 14.7' inlay is included to indicate the position of JD2 which is $\sim1.38$ deg North-North-West of NGC~2403 near the dwarf galaxy DDO~44. At 3.2 Mpc, 1' = 0.93 kpc. We find D2 and JD2 to be foreground stars. 
    }
    \label{pos}
\end{figure}

\begin{table}
	\centering
	\caption{NGC 2403 Globular Cluster Candidate Spectroscopic Observations}
	\label{tab:table1}
	\begin{tabular}{lccccl} 
		\hline
		ID & RA & Dec & RV & S/N & Notes\\
		& (2000) & (J2000) & (km/s) & &\\
		\hline
		C1 & 07:36:10.40 & +65:39:35.0 & 72 & 58 & Disk\\
		C4 & 07:36:04.43 & +65:33:59.8 & 158 & 62 & Disk\\
		D2 & 07:37:26.99 & +65:28:20.9 & 16 & 64  & MW Star\\
		D6 & 07:35:05.74 & +65:45:27.4 & 165 & 22 & Halo\\
		F1 & 07:35:12.43 & +65:43:15.8 & 210 & 45 & Halo\\
		F16 & 07:38:01.52 & +65:33:00.6 & 165 & 43 & Disk\\
		JD1 & 07:37:52.89 & +65:34:07.6 & 145 & 46 & Halo\\
		JD2 & 07:33:34.58 &	+66:56:29.8 & 0 & 12 & MW Star\\
		JC15 & 07:36:18.55 & +65:38:33.0 & 96 & 8 & Disk\\
		\hline
		F46 & 07:36:29.16 & +65:40:33.5 & 140 & -- & Halo\\ 
		\hline
	\end{tabular}
	\\
	Notes: Radial velocity and S/N per \AA~ are measured from our KCWI spectra. The radial velocity of F46 was measured from a  HIRES spectrum by L21. Typical velocities uncertainties are on the order of 10 km s$^{-1}$.  
\end{table}

\subsection{New Candidates} 


We used the characteristics of the candidates from
\citet{1984A&A...130..162B} and \citet{2007ApJ...664..820D} to help select additional GC candidates for spectroscopic follow-up. 
In particular, using SDSS DR7 we searched for slightly resolved objects with $r$-band Petrosian radius $< 2$~arcsec, $i$-band magnitude in the range 17 $<$ $i$ $<$ 19, colour in the range 0.7 $<$ $g-i$ $<$ 1.1 and 2.0 $<$ $u-z$ $<$ 3.5. Two of these candidates were targeted for follow-up spectroscopy, which are named JD1 and JD2. 

In the case of JD2, it was not selected to be a 
GC candidate of NGC 2403 as it lies at a projected distance of $\sim$75 kpc (and hence greater in physical distance), but it does lie near the dwarf satellite galaxy DDO~44 and might be associated with it (given its mass, a couple of GCs might be expected).


We selected further candidates from imaging data obtained with Hyper Suprime-Cam (HSC) on the Subaru 8.2m telescope. These data were gathered as part of the Magellanic Analogs Dwarf Companions and Stellar Halos (MADCASH) survey, a deep, wide-area search for dwarf satellites of nearby Magellanic Cloud analogs. In addition to the deep exposures of NGC~2403 first presented in \citet{2016ApJ...828L...5C}, 
short (30s) exposures on each field were also taken to prevent bright GC candidates from saturating. Here we use these g and i band images taken on 2016/2/9, with seeing of $0.89\arcsec$ and $0.59\arcsec$, respectively. 
After combining the g and i band source catalogs, we matched them to Gaia EDR3 \citep{2016A&A...595A...1G, 2021A&A...649A...1G}, selecting the best match within $1\arcsec$ of each source. 
We then selected all objects with magnitudes $17 < i < 20.5$, colours $0.5 < (g-i) < 1.3$, and at least $4\arcmin$ from the centre of NGC~2403 (to avoid the most crowded regions of the images). We further removed all objects whose Gaia proper motions or parallaxes are measured with $>3\sigma$ significance, as GCs at $\sim3$~Mpc distances should not have measurable proper motions or parallaxes.

After these initial selection criteria, we explored aperture magnitudes from the HSC data, and quality parameters from Gaia that are known to efficiently select extended objects. Because we have a sample of known NGC~2403 GCs, we can confirm that any selection criteria we implement effectively select known GCs. Figure~\ref{fig:i3_6} shows the difference between the HSC i band magnitudes measured in apertures of 3-pixel and 12-pixel radii, which we label $i_3 - i_{12}$, as a function of PSF magnitude. Objects that are very extended (i.e., galaxies) will have a lot of flux beyond their 3-pixel radius, and thus have high values of $i_3 - i_{12}$, while stars should have a small amount of flux beyond a 3-pixel radius, which will be primarily determined by the extendedness of the PSF (and thus roughly the same for all stars). The locus of stars is at $i_3 - i_{12} \sim 0.35$ in Figure~\ref{fig:i3_6}. Extended galaxies will be well above this locus, while GCs, which are only slightly extended, will have intermediate flux ratios in this figure. 

We note that 3 candidates do not appear in 
Figure~\ref{fig:i3_6}. They are: JD1 which is undetected because it lies in a crowded region of the NGC 2403 disk, JD2 which is beyond the HSC field of view and F1 as it does not have reliable measurements from Gaia (it nonetheless appears extended according to its measured $i_3 - i_{12}$ value). Candidate D2 has flux ratios in this figure consistent with being a point source, and also a large measured proper motion  
and is thus likely a foreground star. 

Figure~\ref{fig:gaia_excess} displays the Gaia parameters ``astrometric excess noise'' (AEN) and ``BP/RP excess'', which have been shown by \citet{2021ApJ...914...16H} to effectively select objects around Cen A that are slightly more extended than the PSF (i.e., candidate GCs), as a function of Gaia G magnitude. We overlay and adopt the curves from Equations~1 and 2 from Hughes et al. to separate candidate extended objects from point-like sources.

While this selection yields two dozen candidates, upon visual inspection, we found many of these are obvious background galaxies. Additional work is therefore required to create a comprehensive list, over the entire luminosity function and spatially, of the 50 plus GC candidates expected around NGC 2403. 

 For the best candidate, which we name JC15, we have obtained a spectrum (see below).  
JC15 lies close to the confirmed GCs D6 and F16 in the selection figures. Most importantly, JC15 shows signs of extendedness in $i_3 - i_{12}$, but does not have extremely large flux ratios. It appears round in the i band HSC images, with some hints of semi-resolved faint stars at the outskirts. 

\subsection{Candidates for Follow-up Spectroscopy}

Our final sample of GC candidates for which we obtain spectra are given in Table 1. It lists the g band magnitude and $g--i$ colour taken from SDSS or HSC imaging. The quoted colours are not corrected for Galactic extinction nor internal extinction within NGC 2403. Based on their projected positions on the galaxy disk, JC15 may experience the most internal extinction (it is the reddest object) and D6 the least. 
We also include in Table 1 (and subsequent tables) the confirmed GC F46. This  \citet{1984A&A...130..162B} GC candidate, was subsequently confirmed by 
\citet{2021arXiv211200081L} (hereafter L21) using the HIRES instrument at the Keck Observatory. 

Fig.~\ref{pos} shows an image of NGC 2403 with the locations of our GC candidates indicated. Other than JD2, all of our GC candidates follow a flattened disk structure along the major axis. Given the lack of a bulge in NGC 2403, we expect confirmed GCs to be associated with either the disk or the inner halo. In Fig.~\ref{montage} we show a montage of SDSS, or if available HST, `postage stamp' images of each GC candidate (including F46 by L21).  

\section{Data}

Globular cluster candidates were observed using the KCWI imager slicer 
\citep{Morrissey2018} 
on the Keck II telescope over several different nights under programs W140, U216, W024 and N195. In particular, they are D2 (2019/10/29), 
C4, D6, JD2 (2020/2/16), JC15 (2021/10/4), and F1, F16, JD1, C1 (2022/1/28).

Exposure times ranged from 10 to 40 mins under 1--2" seeing conditions. We used the medium field-of-view and the BH3 grating (with a spectral resolution R = 9000) for C4, D2, D6, JD2. For C1, F1, F16 and JD1 we used the large field-of-view and the BL grating (R = 900). Standard stars were observed in the same setups as the target objects. Some lower S/N spectra were also obtained for some of these candidates but they are not used in this current work.


\begin{figure}
	\includegraphics[width=1.0\columnwidth]{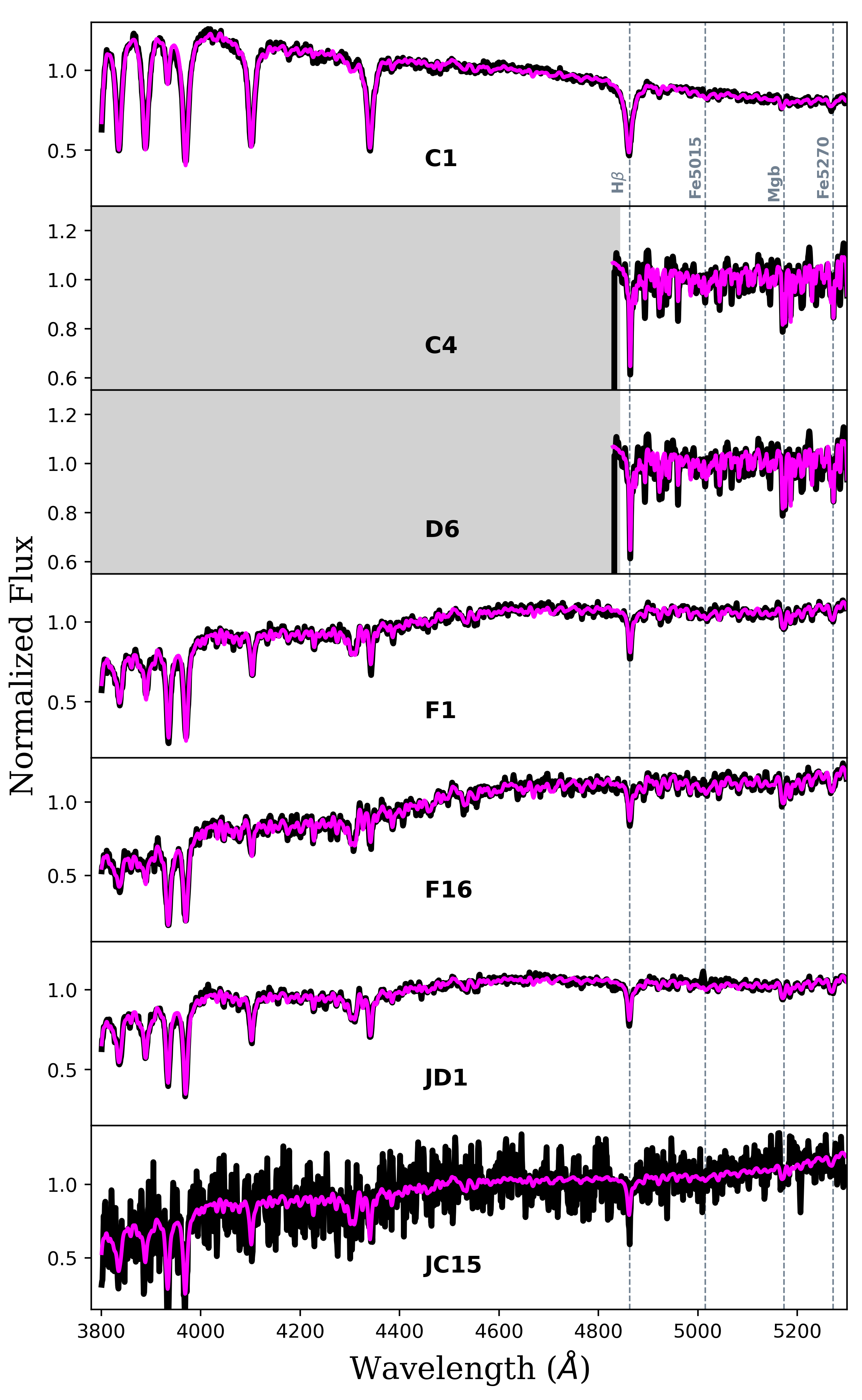}
    \caption{KCWI spectra for each confirmed globular cluster as labelled. Original spectra, shifted to rest wavelength, are shown in black and 
    our model fits in magenta. The key absorption lines used for the analysis are shown in vertical lines (e.g., H$\beta$, Fe5015, Mgb and Fe5270), as labelled in the top panel. Note that different KCWI grating setups were used in the observations, and thus the grey area corresponds to wavelengths not covered by the shorter grating, the BH3}.
    \label{spectra}
\end{figure}

The data were processed using the standard KCWI data reduction pipeline. We took the non-sky subtracted, standard star calibrated cubes and cropped them to the wavelength range common to all slices to use them for the remainder of our analysis. The globular clusters were sky subtracted by taking an appropriately sized region centred on each cluster and subtracting off an 
on-chip offset region as `sky'. Where multiple exposures were taken on the same object these were mean stacked to produce a single spectrum for each GC. Final integration times for each object were: 1440s for C1, 2400s for C4, 1200s for D02, 1200s for D06, 1200s for F1, 1440s for F16, 
1200s for JD1, 900s for JD2 and 900s for JC15. 

Table 2 lists the candidate coordinates, signal-to-noise ratio of each spectrum and our measured radial velocity (with a barycentric correction applied) with typical uncertainties of $\sim$10 km s$^{-1}$. All of the measured radial velocities are consistent with those expected from NGC 2403, however they are also within the velocity range of Milky Way stars.

The GC candidates shown in Fig.~\ref{pos} are colour coded according to their measured radial velocity (RV).
The general rotation of the stellar and HI disk  from \citet{2002AJ....123.3124F} 
shows that the SE side of NGC~2403 is receding at 250 km $^{-1}$ while the NW side is `approaching' us at 10 km s$^{-1}$ (the maximum disk rotation velocity is around 130 km s$^{-1}$). The mean RV of NGC~2403 is 133 km s$^{-1}$. 
On the basis of measured velocities relative to the galaxy's velocity field we suggest that D6, F1 and JD1 (and F46) are GCs likely associated with the halo rather than the disk of NGC 2403. Given these GCs all lie within a projected distance of 13 kpc they may be more akin to {\it inner} halo GCs as seen in the Milky Way's GC system. 
The objects consistent with the general disk rotation are C1, C4, F16 and JC15.

Although the dwarf galaxy DDO~44 reveals a tidal stream due to likely interaction with NGC 2403 \citep{2019ApJ...886..109C}
it has a recession velocity of 255 km s$^{-1}$ and so JD2 with a velocity of $\sim$0 km s$^{-1}$ is very unlikely to be 
associated with DDO~44. We conclude that JD2 is a likely foreground star. 
Candidate D2 has a blue colour, a near zero radial velocity, a similar flux in 3 and 12 pixels in our HSC imaging and Gaia parameters consistent with a Milky Way star. So despite the extended size for D2 given by \citet{2007ApJ...664..820D}, 
we conclude that it is actually a foreground star. 

The status of each GC candidate is summarised in Table 2, with 7 being confirmed GCs and 2 as foreground stars. We will therefore no longer consider these stars further. 
Fig.~\ref{spectra} shows the KCWI spectra for the final sample of 7 confirmed GCs (and highlights the different wavelength scales used for the observations). The key absorption lines are indicated.

\section{Stellar Population Analysis}

\begin{table}
	\centering
	\caption{NGC 2403 Globular Cluster Stellar Populations}
	\label{tab:table2}
	\begin{tabular}{lcccc} 
		\hline
		ID & Age & [Z/H] & [Fe/H] & M$_{\ast}$\\
				& (Gyr) & (dex) & (dex)  & (10$^5$ M$_{\odot}$)\\
		\hline
		C1 & 0.4$\pm$0.2 & --0.61$\pm$0.29 & --0.4$\pm$0.50 & 0.5\\
		C4 & 12.5$\pm$1.6 & --2.07$\pm$0.18 & --2.20$\pm$0.15 & 12\\
		D6 & 7.8$\pm$1.5 & --1.25$\pm$0.04 & --1.20$\pm$0.15 & 2.8\\
		F1 &  9.0$\pm$2.9 & --1.25$\pm$0.20 & --1.35$\pm$0.30 & 5.0\\
		F16 &  9.3$\pm$2.2 & --0.71$\pm$0.29 & --1.15$\pm$0.20 & 4.1\\
		JD1 & 8.6$\pm$2.9 & --1.24$\pm$0.31 & --1.60$\pm$0.30 & 4.5\\
		JC15 & 10.8$\pm$2.1 & --1.41$\pm$0.28 & --1.50$\pm$0.20 & 2.9\\
		\hline
		F46 &  $\le$13 & -- & 
		--1.70$\pm$0.03 & 11\\ 
		\hline
	\end{tabular}
	\\
    Notes: See text for details of possible systematic effects on stellar population parameters. F46 measured from a HIRES spectrum by L21. 
\end{table}

To analyse the stellar populations of the 7 GCs, we follow the method outlined in \citet{2021MNRAS.503.5455F} for compact stellar systems. Briefly, we use the MILES Single Stellar Population (SSP) library 
\citep{2010MNRAS.404.1639V} models with the BaSTI
isochrones, considering templates that range from metallicities of
[Z/H] = --2.42 to +0.40 dex and that cover ages from 0.03 to 14 Gyr, assuming a universal Kroupa IMF. The SSP models include different alpha element ratios, from solar to super-solar \citep{2016MNRAS.463.3409V}. 
The work of \citet{2007MNRAS.379.1618M} compared the SSP models of Vazdekis et al. to other SSP models and found good consistency for Milky Way GCs. 
We run pPXF using multiplicative polynomials and applying a regularization value that ensures that the resulting star formation history is the smoothest possible while
maintaining a realistic fit. The best fit SSP model using pPXF 
\citep{2012ascl.soft10002C} for each GC is overplotted in Fig.~\ref{spectra}. For the results, we adopt the means of the mass-weighted stellar population parameters. However, for the [Fe/H] metallicity that is measured directly from the absorption line indices. For the [$\alpha$/Fe] ratios we employ the Mg$_b$ and Fe5015 lines (which are common to all spectra;  these same pairs were employed by the SAURON survey of early-type galaxies; \citet{2007MNRAS.379..445P}). We find that only JD1 has a clear super-solar alpha element enhancement and, in this case we employ the corresponding alpha-enhanced, rather than base, MILES models. Otherwise the GCs are consistent with a solar alpha element ratio.

Table 3 presents the mean mass-weighted stellar populations that we measure for the ages and metallicities with uncertainties. Quoted uncertainties are determined using a Monte Carlo technique in which we have varied different ingredients in the pPXF analysis, (e.g. polynomials, degree, regularization, alpha vs. base models) to determine the overall uncertainties. Stellar masses are calculated from $g$-band magnitudes and the resulting mass-to-light ratio from the SSP analysis.  We note the very young age of the cluster C1 (400 Myr) and the strong Balmer absorption lines in its spectrum. Our SSP analysis reveals no evidence for a hidden old stellar population for this GC. Although we continue to refer to C1 as a GC, some prefer the `young massive cluster' nomenclature.

For F46, we list the stellar population parameters determined independently by 
\citet{2021arXiv211200081L}, i.e. their assumed age and measured [Fe/H] metallicity (they found [$\alpha$/Fe] = +0.15). 
Their assumed age, of 13 Gyr, can be considered an upper limit. For the stellar mass we assume the same ratio as JC15 (i.e. M/L$_g$ $\sim$ 2) which gives a stellar mass of 11 $\times$ 10$^5$ M$_{\odot}$ in reasonable agreement with the dynamical mass calculated by L21 of 15 $\times$ 10$^5$ M$_{\odot}$. 

\begin{figure}
	\includegraphics[width=0.8\columnwidth,angle=-90] {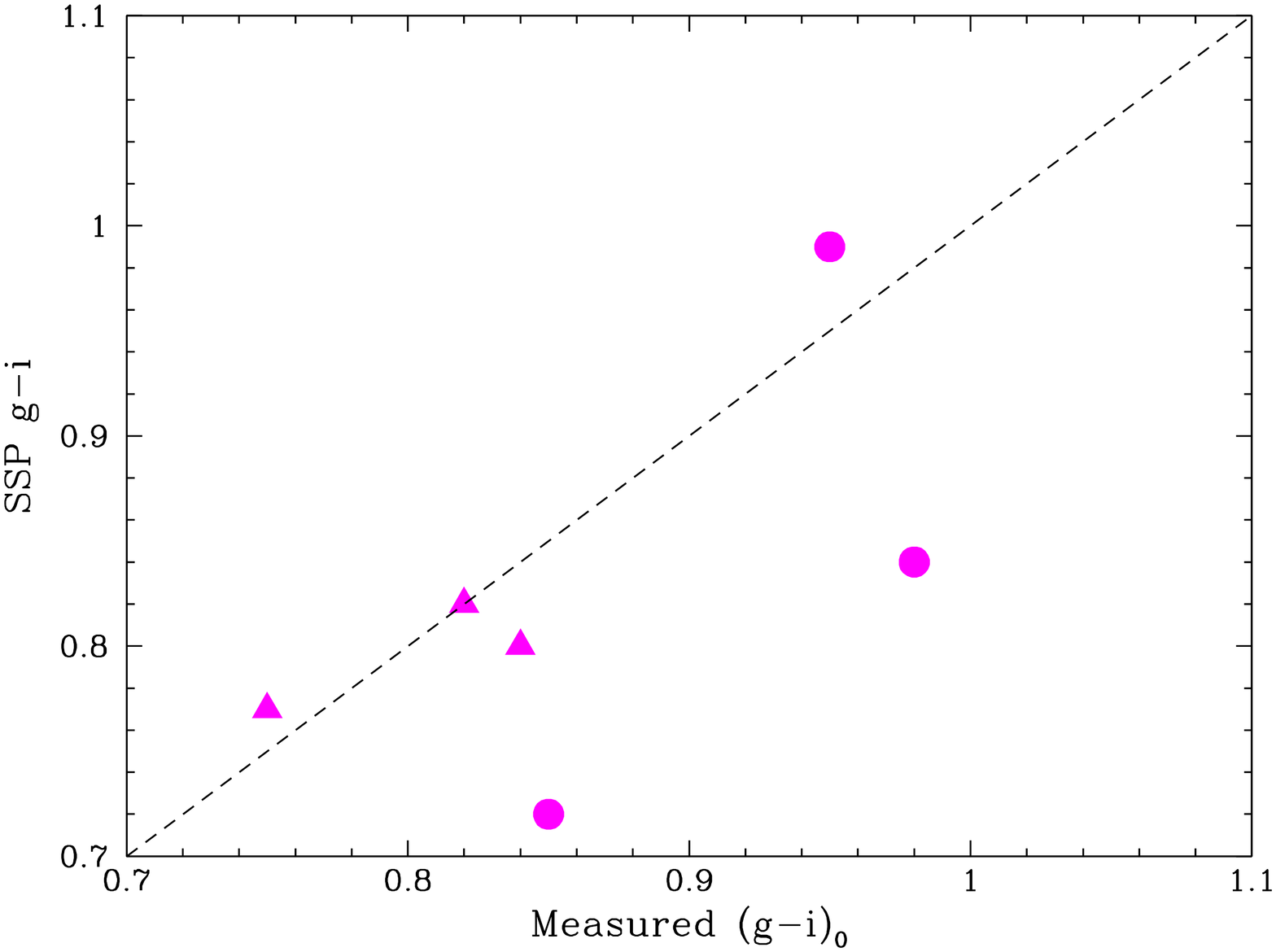}
   \caption{Comparison of $g-i$ colours predicted from our single stellar population (SSP) analysis vs the observed colours (corrected for 
   Galactic extinction) for NGC 2403 globular clusters. Magenta filled circles show disk GCs and filled triangles show inner halo GCs. The plot focuses on the old, red GCs: the young blue cluster C1 (with SSP $g-i$ of 0.2 and observed $g-i$ of 0.15) is not shown.  
   The dashed line shows a one-to-one line. Corrections for internal extinction would move data to bluer observed colours. 
   } 
    \label{colours}
\end{figure}

Using the stellar parameters determined from our SSP analysis, we predict $g-i$ colours for our GCs. These predicted colours are compared to the measured colours, corrected for Galactic extinction (A$_V$ = 0.11), in Fig.~\ref{colours}. 
The predicted and Galactic extinction-corrected colours are in reasonable agreement, thus providing further confidence in our stellar population analysis. Correction for internal dust extinction within NGC 2403 for the disk GCs could improve the agreement further.

Before discussing our stellar population results it is worth reviewing the potential biases and caveats in the object selection and data analysis. Both the literature and our own selection of GC candidates are naturally biased away from the inner disk due to the possible confusion with young star clusters and HII regions. This might tend to bias our sample against the selection of metal-rich disk GCs. 

The stellar masses calculated for our sample of GCs (see Table 3) are all but one above 10$^5$M$_{\odot}$. With a mean GC mass of around 2$\times$10$^5$M$_{\odot}$ for the Milky Way's GC system, this indicates that our sample is drawn from the most massive or brightest (for comparison the GC turnover magnitude is M$_g$ $\sim$ --7.2) of the GCs in NGC 2403. We note that selecting the more massive GCs from the Milky Way's system would create a bias towards selecting accreted GCs \citep{2005MNRAS.360..631M}.

There is the possibility of contamination in our GC spectra by emission lines associated with star formation in the disk of NGC 2403. If present, H$\beta$ emission would tend to fill in the H$\beta$ absorption line resulting in an older inferred age if not corrected for. If contamination from emission lines was present, we would expect to see the [OIII] line at 5007~\AA~ -- a visual inspection reveals no clear evidence for [OIII]. Furthermore, there is no evidence for emission lines from our pPXF analysis of the entire spectrum. Thus 
we conclude that our spectra do not suffer from such contamination. 


Moreover, the observed GCs may contain blue horizontal branch (BHB) stars. These hot stars can give the impression of a young spectral age while having limited effect on the derived metallcitity and alpha elements \citep{2018ApJ...854..139C}. The study of \citet{2007MNRAS.379.1618M} examined the effects of BHB stars on the derived stellar population parameters from integrated spectra and found a limited impact when a large spectral range was used. We thus consider that the ages of those GCs with a shorter baseline should be considered lower limits, as without higher quality and/or longer baseline spectra we are unable to rule out the presence of BHB stars.

\section{Results and Discussion}

In this work we obtain spectra, radial velocities, stellar population parameters (age, metallicity and alpha abundances) and stellar masses for 7 confirmed GCs associated with NGC 2403. One additional GC, F46, was previously observed using HIRES at the Keck Observatory by \citet{2021arXiv211200081L} and we include it in our subsequent analysis. They derived a radial velocity (confirming association with NGC 2403), a low metallicity and a mildly enhanced alpha element ratio. Previously, \citet{1984A&A...130..162B} managed to obtain low quality spectra of 3 GC candidates (C3, C4 and F21). For C3 they noted similarities to a A-type spectrum indicating an age of around $\le$1 Gyr. For the only object in common with this study, C4, they described the spectrum as being that of a classical GC (which we confirm). For F21 they concluded it was a background galaxy. 

\begin{figure*}
\includegraphics[width=5in,angle=-90]{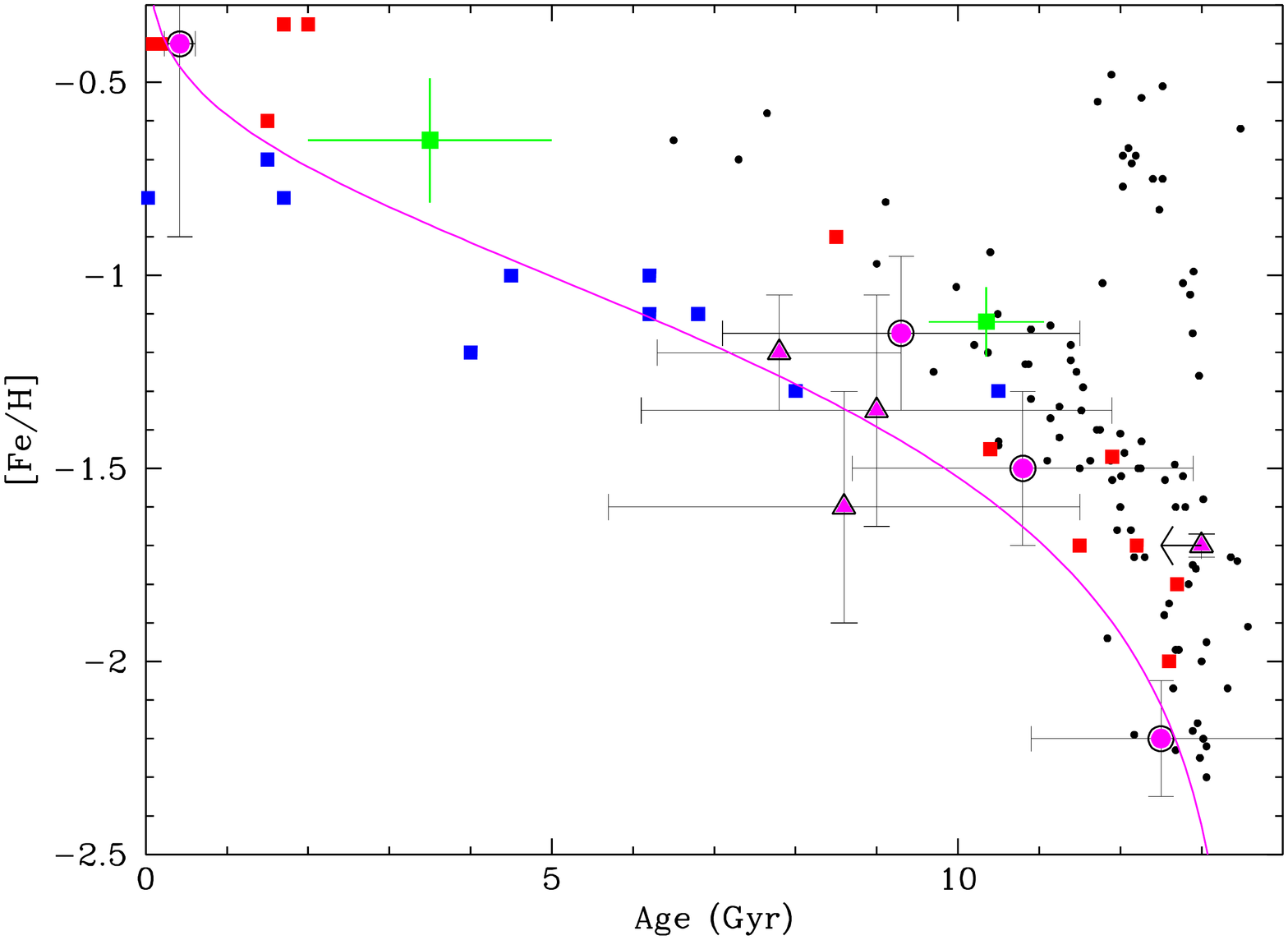}
    \caption{Age-metallicity relation for NGC 2403 globular clusters. Magenta filled circles show NGC 2403 disk GCs and filled triangles show inner halo GCs. 
    For GC F46 an age of 13 Gyr was assumed in L21.  The green squares with error bars show an intermediate-age GC and the mean value for several outer halo GCs in M33. Red and blue squares (without error bars) show a selection of LMC and SMC GCs respectively, with reliable ages and metallicites. Small black circles show Milky Way GCs. 
The magenta line shows a representative chemical enrichment model for the NGC 2403 GCs.
    The NGC 2403 GCs are more metal-poor on average at a given age than those in the Milky Way but similar to those in the SMC. 
    }
    \label{amr}
\end{figure*}

The stellar populations, and possible biases/caveats, for the Keck-observed GCs are summarised in Section 4 and Table 3. 
Bearing in mind that 8 GCs out of a possible GC system of around 50 may not be representative, we discuss our sample of high-mass disk/halo GCs in the context of GC systems in other nearby galaxies below.

Our data indicate that the GC system of NGC 2403 was formed over an extended period of time from early in the Universe until today. The GCs have near solar alpha-element ratios with one presenting super-solar enhancement. Compared to the two subpopulations of the Milky Way's GC system, the old GCs studied here are all relatively metal-poor ([Fe/H] $\le$ --1); only the very young cluster C1 is metal-rich. We note that \citet{2003AJ....125.3046D} also measured a low metallicity of [Fe/H] = --2.2 for field stars in the outer halo (projected $\sim$22 kpc along the minor axis, beyond our most distant GC) of NGC 2403. This metallicity is comparable to those for the oldest GCs in our sample.

In Fig.~\ref{amr} we show the ages and [Fe/H] metallicities  for our NGC 2403 GCs, along with those of the Milky Way, the Magellanic Clouds and M33 (a possible analogue galaxy to NGC 2403). 
For M33 we show a confirmed intermediate-aged (2--5 Gyr) GC from \citep{2006ApJ...646L.107C} which has a metallicity of [Fe/H] = --0.65$\pm$0.16  \citep{2000AJ....120.2437S}. We also show the mean values for a dozen old M33 GCs (i.e. age = 10.35$\pm$0.71 Gyr and [Fe/H] = --1.12$\pm$0.09) from \citet{2015MNRAS.451.3400B}. We refer the reader to \citet{2021arXiv211200081L} for arguments that some of the Beasley et al. objects may actually be foreground stars (their exclusion does not significantly change the mean values). For the Magellanic Clouds we take the reliable ages and metallicities from the appendix of \citet{2021MNRAS.500.4768H}, i.e. those with a confidence code of 1. 
We take ages and metallicities for Milky Way GCs from \citet{2020MNRAS.493..847F}. 
In the Milky Way, the in-situ formed bulge and disk GCs formed over a short time period (1-2 Gyr) at early times ($\sim$ 13 Gyr ago), whereas the younger GCs are all the result of ex-situ formation and subsequent accretion of their host dwarf galaxy  \citep{2020MNRAS.493..847F}.

Our observed NGC 2403 GCs in Fig.~\ref{amr} are coded by whether we have assigned them to the halo or to the disk of NGC 2403 on the basis of their relative velocities (see Table 2). We find no clear distinction between disk and halo GCs in terms of their average age or metallicity. 
The (disk) GC F46 is shown with an upper age limit of 13 Gyr as assigned by L21. 
Given that NGC 2043 has little or no bulge (similar to M33), we do not expect bulge GCs to be present, as they are in the Milky Way. 
At a given age, the NGC 2403 GCs observed in this study have metallicities most similar to those of the SMC, slightly less than those of M33 and the LMC, and 
systematically lower than GCs accreted from dwarf galaxies onto the Milky Way.
In terms of the age distribution, the Magellanic Clouds, M33 and NGC 2403 all appear to have formed star clusters from early times ($\sim$ 13 Gyr ago) until just a few Gyr ago. We remind the reader of C3 in NGC 2403, which \citet{1984A&A...130..162B} claimed it has an A-type spectrum indicative of a $\le$1 Gyr old GC. Furthermore massive ($\sim10^5$ M$_{\odot}$),
young ($\sim$0.1 Gyr) star clusters have been identified by \citet{1999A&A...345...59L} to be forming up to the present day in NGC 2403.



To further discuss the age-metallicity distribution of our observed GCs, we also show in Fig.~\ref{amr} the age-metallicity relation (AMR) for a leaky-box chemical enrichment model. The chemical enrichment models assume non-enriched gas 13.5 Gyr ago and an  effective yield of $p$ (i.e. elements returned to the ISM for subsequent star formation after accounting for outflows of enriched material), which is related to stellar metallicity by the equation:\\

[Fe/H] = - $p$ ln ($t$/13.5) \\

\noindent
Such a model provides a good representation for the GCs from disrupted satellites in the Milky Way. Indeed, this aspect of dwarf galaxy GC systems was used by \citet{2019MNRAS.486.3134K} and \citet{2020MNRAS.493..847F} to associate individual Milky Way GCs with disrupted satellites. The yields for the five largest disrupted satellites, including the Sgr dwarf, were found to range from 0.22 to 0.35. The model shown in Fig.~\ref{amr} has $p$ = 0.1 and is very similar to the one derived from the stars in the SMC (M$_{\ast}$ = 0.4 $\times$ 10$^9$ M$_{\odot}$)
from \citet{2013MNRAS.436..122L}. 


The steepness of an AMR (i.e. the metallicity at a given age), as represented by the yield, is primarily driven by the mass of the host galaxy. Given that the AMR for NGC 2403 appears to be lower than that of dwarf galaxies accreted onto the Milky Way and perhaps also the GCs of M33 (which has a similar luminosity to NGC 2403), other factors could be at play. 
One factor that may act to lower an AMR is the outlfow of enriched material. In a study of nearby spirals and dwarf galaxies, \citet{2007ApJ...658..941D} concluded that {\it "Metal-enriched outflow is therefore the only viable mechanism for producing galaxies with low effective yields."} The presence of an extended `HI halo' around NGC 2403 led \citet{2002AJ....123.3124F} 
to suggest that it may be the signature of a galactic fountain. This outflow from the disk to the halo may contribute to the low effective yield as argued by \citet{2007ApJ...658..941D}. Furthermore, there is a step change in the effective yield to lower values at rotation speeds of V$_{rot} \le 150$ km s$^{-1}$ \citep{2002ApJ...581.1019G}. NGC 2043, with V$_{rot}$ $\sim$ 130 km s$^{-1}$ falls close to this key transition.

Cosmological simulations predict that galaxies of NGC 2403's mass have not undergone significant accretion (e.g. in the model of \citet{2019MNRAS.488.5409C}) and only in-situ formed GCs are expected. Indeed there are several arguments against NGC 2403 having experienced a massive merger event in the past, which would mean few, if any, accreted GCs.
NGC 2403 is relatively isolated and may only be infalling into the low mass M81 group for the first time (there is no evidence of ram pressure stripping). It also reveals a remarkably uniform regularly rotating HI disk \citep{2008AJ....136.2648D}. Its old stellar disk has a uniform metallicity of [Fe/H] $\sim$ --1 (consistent with our most metal-rich GCs), indicating that it is well-mixed, and has an unbroken surface brightness profile \citep{2013ApJ...765..120W}. This suggests a relatively undisturbed galaxy despite its  probable interaction with the dwarf satellite DDO~44 \citep{2021ApJ...909..211C}. 
\citet{2003AJ....125.3046D} found evidence for AGB stars (ages $\le$ 1 Gyr) beyond the disk of NGC 2403 and argued that they formed in-situ and not from accretion. We also note that F46 does not have the anomalous chemical signature indicative of an accreted GC \citep{2021arXiv211200081L}. While we cannot rule out accreted GCs, our sample of GCs is likely formed, over many Gyr, from in-situ gas.

\section{Conclusions}

In this work we studied a sample of relatively high mass globular clusters (GCs) associated with the low mass, bulgeless galaxy NGC 2403. 
Based on GC candidates in the literature, and new candidates selected from SDSS and our Hyper Suprime-Cam imaging, we obtained spectra of nine candidates using the KCWI instrument at the Keck Observatory. We confirm seven of them to be GCs associated with NGC 2403, with the remaining two likely to be foreground stars. Our results are supplemented with one additional GC observed using HIRES, also from Keck, by \citet{2021arXiv211200081L}. 

By comparing the radial velocities of the eight confirmed GCs with the overall rotation of the HI disk, we assign four GCs to the disk and four to the halo. Single stellar population model fits indicate that the GCs were formed over many Gyr (with the caveat that the presence blue horizontal branch stars can make GCs appear much younger than they are). We find most to be consistent with solar alpha element ratio. However, the old GCs are all found to be metal-poor, i.e. [Fe/H] $\le$ --1. The one young, massive cluster we obtain a spectrum for is more metal-rich. Cosmological simulations predict that few, if any,  accreted GCs will be found in galaxies of the mass of NGC 2403.
Based on the lack of disturbance in the stellar and HI disk, and its relative isolation, both the disk and halo GCs of NGC 2403 likely formed over an extended period from in-situ gas.

Our NGC 2403 GCs are systematically lower in metallicity at a given age compared to the Milky Way's GC system and perhaps also to the GCs in M33. However, they exhibit a similar age-metallicity relation to the GCs of the Small Magellanic Cloud. 
The age-metallicity relation of the NGC 2403 GCs can be approximated by a leaky-box chemical enrichment model with a similar effective yield than that inferred for the SMC stars. Outflows of enriched material may explain the relatively low yield inferred for NGC 2403's GC system.

\section*{Acknowledgements}

We thank. S. Larsen for supplying information regarding the globular cluster F46 prior to publication and for his comments on our paper. We thank the MADCASH team for the original 
HSC observations and D. Sand for his comments.
We thank the referee for several suggestions that have improved this paper. 
AFM has received support from the Severo Ochoa Excellence scheme (CEX2019-000920-S) and through the Postdoctoral Junior Leader Fellowship Programme from 'La Caixa' Banking Foundation (LCF/BQ/LI18/11630007).
AJR was supported by National Science Foundation grant AST-1616710 and as a Research Corporation for Science Advancement Cottrell Scholar. JLC acknowledges support from National Science Foundation grant AST-1816196.

This work was supported by a NASA Keck PI Data Award, administered by the NASA Exoplanet Science Institute. Data presented herein were obtained at the W. M. Keck Observatory from telescope time allocated to the National Aeronautics and Space Administration through the agency's scientific partnership with the California Institute of Technology and the University of California. The Observatory was made possible by the generous financial support of the W. M. Keck Foundation.
The authors wish to recognize and acknowledge the very significant cultural role and reverence that the summit of Maunakea has always had within the indigenous Hawaiian community. We are most fortunate to have the opportunity to conduct observations from this mountain.

\section*{Data Availability}

Keck data are available from the Keck Observatory Archive (KOA) at https://www2.keck.hawaii.edu/koa/public/koa.php. 
Sloan Digital Sky Survey (SDSS) photometry is available at
http://skyserver.sdss.org/dr7/en/home.aspx. 
HST data are available from the MAST archive at 
https://archive.stsci.edu/missions-and-data/hst.



\bibliographystyle{mnras}
\bibliography{N2403} 




\appendix

\section{Globular cluster candidate imaging}

Images, from HST or SDSS, of our globular cluster candidates studied in this work, along with the confirmed GC (F46) of \citet{2021arXiv211200081L}. 
We find two candidates (D2 and JD2) to be foreground stars.

\begin{figure}
	\includegraphics[width=0.95\columnwidth,angle=0]{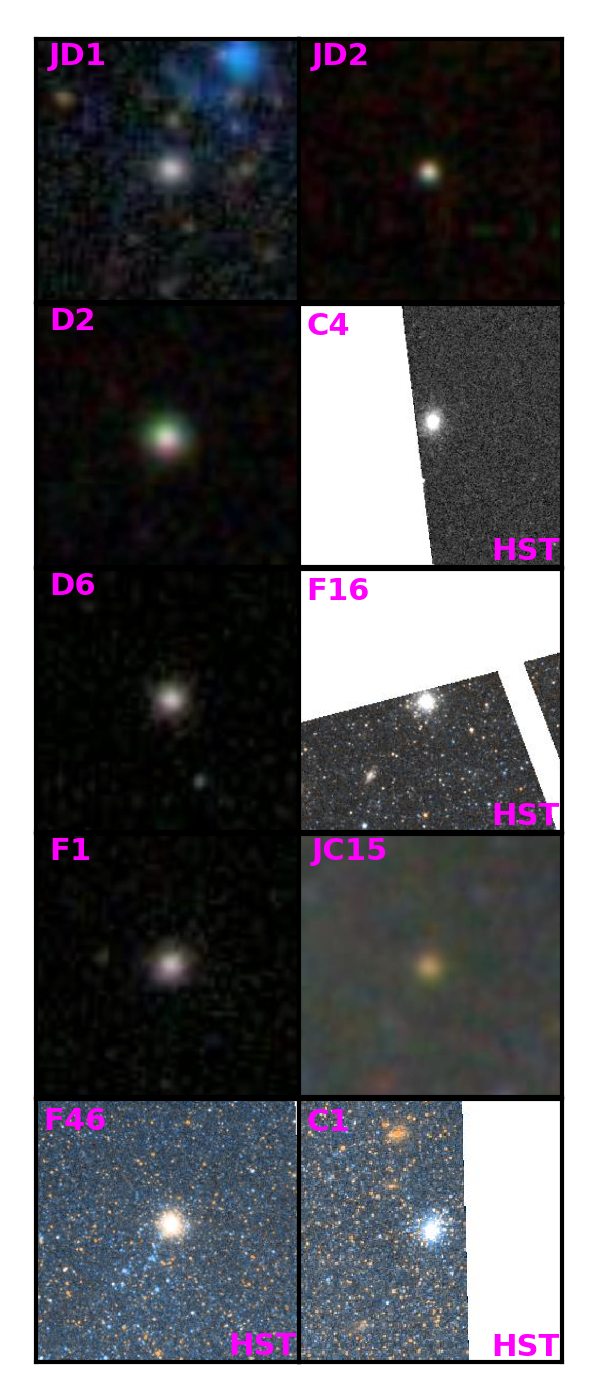}
    \caption{A montage of the globular cluster candidates studied in this work taken from SDSS 
    or, if available, from HST imaging.  Globular cluster IDs are labelled. Each image postage stamp covers a field of view of 25.6" $\times$ 25.6". At 3.2 Mpc, 1" = 15 pc. We conclude that D2 and JD2 are foreground stars (see text for details). 
    }
    \label{montage}
\end{figure}



\bsp	
\label{lastpage}
\end{document}